\documentclass[12pt]{article}
\usepackage{epsfig}
\usepackage{graphicx}
\usepackage{cite}
\newcommand{\beq}{\begin{equation}}
\newcommand{\eeq}{\end{equation}}

\newcommand{\forget}[1]{}

\newcommand{\turnmeon}[1]{}
\newcommand{\turnmeoff}[1]{#1}

\turnmeoff{

\textwidth 490pt
\flushbottom
\textheight 630pt
\topmargin 14pt 
\headheight 0pt
\headsep 0pt
\footskip 32pt
\oddsidemargin -10mm 
\parindent 7mm
\parskip 0pt

}

\turnmeon{

\textwidth 510pt
\flushbottom
\textheight 650pt
\topmargin 12pt 
\headheight 0pt
\headsep 0pt
\footskip 30pt
\oddsidemargin -10mm 
\parindent 6mm
\parskip 0pt

}

\newcommand  {\PNAS}     {{\it Proc.\ Natl.\ Acad.\ Sci.\ } USA\ }

\newcommand  {\Sci}      {{\it Science\ }}

\begin{document}
\begin{titlepage}
\hfill LU TP 00-18

\hfill May 22, 2000

\begin{center}
{\large \bf Shuffling Yeast Gene Expression Data} 

\vspace{0.9cm}

{\sl Sven Bilke}\footnote{eMail: sven@thep.lu.se} 

\vspace{0.3cm}

Complex Systems Division, Department of Theoretical Physics\\
University of Lund, S\"olvegatan 14A, S-22362 Lund, Sweden\\
{\rm http://www.thep.lu.se/complex/}
\vspace{0.3cm}
\end{center}

\vfill
\turnmeon{
Classification: Biological Sciences, Genetics.

Corresponding author: S. Bilke, tel: +46-46-2220667, 
fax: +46-46-2229686.

Manuscript information: 12 text-pages, 2 figures.

Word and character count: 103 words in the abstract, 25800 
characters in the paper.
\vfill
\end{titlepage}

\newpage
}
{\bf{\centerline{Abstract}}}
\noindent
A new method to sort gene expression patterns into functional groups
is presented.  The method is based on a sorting algorithm using a
non-local similarity score, which takes {\em all} other patterns in
the dataset into account. The method is therefore very robust wih
respect to noise.  Using the expression data for yeast, we extract
information about functional groups.  Without prior knowledge of
parameters the cell cycle regulated genes in yeast can be
identified. Furthermore a second, independent cell clock is
identified. The capability of the algorithm to extract information
about signal flow in the regulatory network underlying the expression
patterns is demonstrated.

\vfill
\turnmeon{\newpage}
\turnmeoff{\end{titlepage}}

\section{Introduction}
The DNA microarray technology \cite{schena95} has greatly facilitated the study
of gene expressions. With a single microarray, the expression of thousands 
of genes can be measured simultaneously. Based on the central dogma it is
reasonable to understand these expression vectors as a description of the
functional state of the cell. The dynamics of the state-trajectory  
observed in expression time series reveals much   information about the 
regulatory network underlying gene expression. A detailed knowledge of this
network would allow for the analysis of  possible states and trajectories, 
including, say, transitions from disease to a healthy state. 

Therefore, it is very tempting to try to infer the regulatory network
from the data. One should, however, be aware of the limitations in 
the data available so far. It is very possible that large  parts of the 
network were inactive for the states observed. These ``unexcited'' parts
of the network can not be deduced from the data. Furthermore important
parts of the regulatory network, like e.g. inter- and intra-cell signalling, 
are observed only indirectly by back reaction on  the gene expression pattern.

Cluster algorithms have been used successfully in the analysis 
of gene expression data. Using for example hierarchical clustering 
it has been demonstrated \cite{esbb98} that many genes, which on biological
grounds are known to be related, are located near by in the  similarity
tree. It is however  difficult to identify genes
which belong to a larger functional context, like for example cell cycle
regulated genes. If two of the corresponding patterns  are expressed with a 
phase difference close to $\pi /4$, they  are uncorrelated and therefore 
placed on remote sites in the similarity tree.

The prior knowledge of the cell cycle frequency $\nu _{\tiny cc}$
was used to identify the cell cycle regulated genes by inspection
\cite{ccwscwwglld98}. In \cite{ssziaebbf98} a spectral filter was used
for this purpose. Expression patterns, for which the spectral energy
at frequency $\nu \approx \nu _{\tiny cc}$ is larger than some threshold,
were selected as cell cycle regulated.
                                           
In this work we use a new method, the re-shuffling algorithm
\cite{b00}, to identify functional groups in the expression data.
The algorithm sorts the data based on a global similarity score, which
makes it very robust with respect to noise. The method does
not distribute the data into clusters. The structure found in the data is 
rather reflected in a re-ordered sequence of expression patterns.
Using this algorithm we are able to identify the cell cycle regulated
genes in  the budding yeast {\em S.~cerevisae} without referring to 
prior knowledge.  Furthermore we find a second, independent clock in the cell. 
The reordered sequence of expression patterns can reflect 
the propagation of a signal in the data. Patterns, which respond to the
same, deformed signal are grouped together, even if they are mutually 
uncorrelated. 

\section{Algorithm}
In this section we briefly describe the algorithm used to analyze the data.
A more detailed description can be found elsewhere \cite{b00}.  

The starting point is a matrix $C_{ij}$ encoding the similarity between
expression patterns $i$ and $j$. This similarity can, for example,  be 
the mutual information or  correlation for the two patterns. 
The purpose of the algorithm is to find a relabeling 
$i \rightarrow \sigma(i)$ such that similar patterns $i,j$, 
i.e. $|C_{ij}| \approx 1$, get similar labels: 
$|\sigma(i) - \sigma(j)| \approx 1$. This is, however, not 
achieved by performing local, mutual, comparisons,  but
rather by letting  expression patterns  move freely in a ``force-field'' 
generated by {\em all} other particles. This field is is described by the 
energy 
\beq
S_{\sigma}(\alpha, \gamma, \lambda) = - \sum _{i,j} 
	\mbox{sgn}(C_{\sigma(i)\sigma(j)}) ^{\gamma}
        |C_{\sigma(i)\sigma(j)}|^{\alpha} 
              \exp  - \frac {d(i, j)^2}{N \lambda}. 
\label{action}
\eeq 
The optimal sorting in the sense described above is the one minimizing this 
energy. The parameter $\alpha $ controls the importance of the similarity
in the sorting procedure, we use $\alpha = 2$.  
The variable $\lambda$ is a localization parameter. For small $\lambda$, 
mainly similarities close in index space contribute
to the energy. This  leads to a local optimization. For large $\lambda $ 
a more global optimization is achieved. The parameter $\gamma$ is used
to switch between maintaining  ($\gamma = 1$) or ignoring ($\gamma = 0$)
the sign of $C_{ij}$. 

Obviously the average distance $\overline{d}$ of indices from all 
other indices is not evenly distributed. It reaches its maximum on the border
$i=1, N$. This non-flat distribution is not desired, therefore we 
use a cyclic distance measure in index space. With
\beq 
d(i,j) = \left \{
\begin{array}{rcl}
 |i - j|     & \quad & {\rm if~} | i - j | < N / 2 \\
 |i - j - N| & \quad & {\rm if~} | i - j | \geq N / 2 
\end{array} \right ., 
\eeq
the first and the last pattern in the list are direct neighbors, the system 
has no boundary and therefore the $\overline{d}$ distribution is flat. 

Unless $C_{ij}$ has a very simple form, the minimization of equation 
(\ref{action}) is a non-trivial task. We use simulated annealing 
\cite{anealing} for this purpose.
\forget{
the 
statistical ensemble of re-labelings $\sigma $described by the partition 
function 
\beq
Z(\alpha, \lambda, T) = \sum _{\{\sigma\}} 
      \exp - \frac{S_{\sigma}(\alpha, \beta)}{T}.
\label{part}
\eeq
}
In the annealing procedure a fictious ensemble temperature $T$ is lowered.
At the beginning, at high temperature, the global aspects of the structure 
contained in  the data should be  built into the order of expression patterns,
while towards the end of the annealing procedure, at low temperature, the more 
local optimization takes place. Therefore the localization parameter
$\lambda $ is lowered together with the temperature from typically 
$\lambda = 1$ to $\lambda = 0.05$ in this procedure.

\forget{
Depending on the complexity of the interaction
$C_{ij}$ the decay into the ground state  may, however, take a very 
long time. The system can be trapped in a meta-stable state,  a local optimum. 
Therefore  an annealing strategy is used to find the global optimum. 
We start with large $T$ and large $\lambda $ to perform optimization
on a large scale and then  lower $T$ and $\lambda $ concurrently.

The statistical system defined by (\ref{part}) can be analyzed numerically
using Monte Carlo methods. A sample of the  equilibrium distribution can be 
obtained as a time series generated by a Markov process: starting from an 
(arbitrary) configuration a sequence of transformations, exchange of
two indices, is applied to the sequence. A proposed transformation
is accepted with the Metropolis probability
\beq 
p(\sigma _1 \rightarrow \sigma_2)  = \left \{ 
\begin{array}{rl}
 1. & {\rm if~} S_{\sigma_2}  < S_{\sigma_1} \\
 \exp \frac{S_{\sigma_1} - S_{\sigma_2}}{T}
 & {\rm if~} S_{\sigma_2}  \geq S_{\sigma_1}
\end{array} \right . .
\eeq
The configuration with the minimal action encountered in this time-series
is used as the optimal solution 
}

\section{Results}
For this work we used  the  expression data for {\em S. cerevisae}, which 
is available on the internet \footnote{\rm http://cellcycle-www.standford.edu}
and described in \cite{ssziaebbf98}. Here we want to demonstrate the 
feasibility  of our algorithm on a subset of this data set. 

The original data consists of $82$ experiments, which were done at different 
time points and/or boundary conditions.  Each experiment provides  
measurements of $6177$ expression ratios. The subset used here was 
extracted in the following way:
\begin{enumerate}
\item Experiments with more than 400 missing expression ratios were 
	removed.
\item Expression patterns (in gene direction) with more than 8 missing 
      ratios were removed.
\item From the remaining  genes measured in $\tau = 69$  experiments we
      kept those $N= 803$ patterns with variance $\sigma > 0.5$.
\end{enumerate}

In the following the Pearson correlation 
\beq
C_{ij} = \frac{ \sum _t ^{\tau} (D_{i,t} - \overline{D_i}) 
                 (D_{j,t} - \overline{D_j}) }
	       { \tau \sqrt{ \mbox{Var}(D_i) \mbox{Var}(D_j)}}
\eeq  
is used as the similarity score in equation (\ref{action}). First the absolute 
value $|C_{ij}|$ is used ($\gamma = 0$), because we are interested in 
analyzing functional groups of genes, which show up  by (anti)-correlated 
expression patterns. The result of the sorting procedure is visualized in 
figure \ref{CorMat}, a graphical representation of the correlation matrix.
In this diagram the intensity of the pixel at coordinate $(i,j)$ 
is proportional to the absolute value $|C_{ij}|$. Red color represents
correlation, green color anti-correlation.
 
\turnmeoff{
\begin{figure}
\centerline{\psfig{file=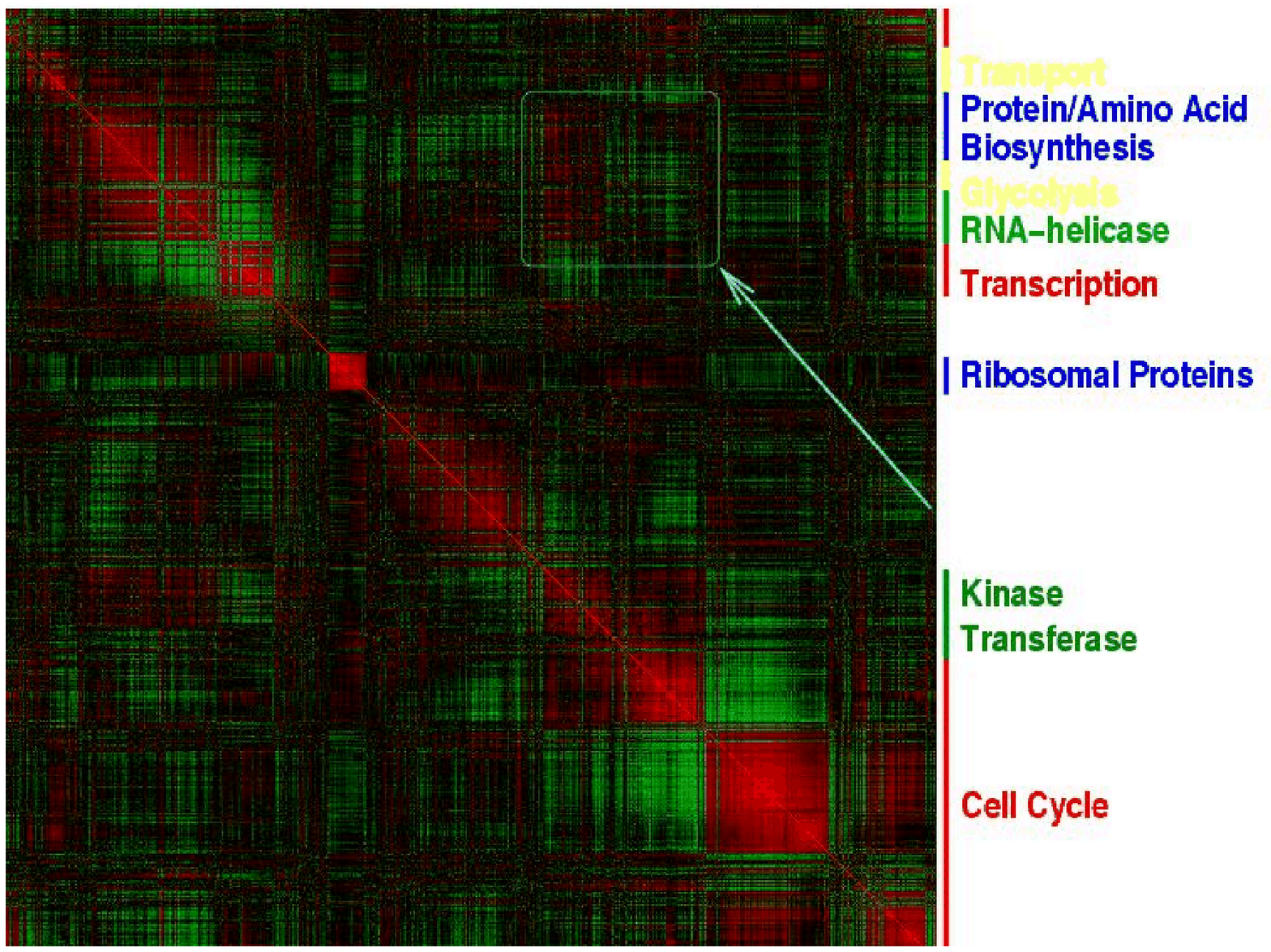,height=7cm}}  
\caption{Correlation Matrix for the 800 most variant gene expression
         patterns. The annotation refers to the most frequent annotation
         found in the database for the genes in this group.}
\label{CorMat}
\end{figure}
}

At coordinates adjacent to the main diagonal of the matrix one 
clearly observes a grouping of gene expression patterns.  
With the help of  the annotations for the genes involved one can verify
that the grouping reflects a classification with respect to gene function. 
The annotation displayed in figure \ref{CorMat} refers to the most frequent 
phrase found in the annotation database for the genes in these groups. 
These typical phrases do not exclusively show up there, their distribution 
over the whole dataset is, however,  strongly peaked in the marked groups. 

The diagram does not only contain information about the dominant
correlation, which  clusters the genes into groups. Sub-dominant 
co-regulations can be seen in the more off-diagonal parts of the matrix. 
As expected, the  non-local interaction (\ref{action}) sorts the groups
on the main diagonal such that clusters with sub-dominant 
co-regulation group together. Therefore the distance of groups on
the main diagonal reflects the relative strength of co-regulation. 
In the off-diagonal correlation coefficients an interesting fine structure
can be  observed: for example in the region marked with an arrow one sees that
two groups, which are internally correlated, can be correlated and 
anti-correlated at the same time. From this observation one can  infer
a fine-structure into the groups on the diagonal.  

The checker board patterns observed in the upper left and lower right 
in figure \ref{CorMat} are very interesting. Obviously they are generated 
by oscillatory processes: adjacent red and green blocks indicate co-regulated,
but mutually exclusive expressed genes. These genes are active in 
different parts of a cycle. By inspection of gene annotations, the
lower right functional group is identified as  cell-cycle regulated 
genes. In \cite{ssziaebbf98} these  genes were
identified using a different method.  Expression patterns, for which
the  Fourier component for frequencies close to the expected cell-cycle
frequency were larger than some threshold were identified as
cell-cycle regulated. For our method, the introduction of a threshold and 
knowledge of the oscillator frequency is not necessary.  

The checker board in the upper 
left corner represents a second cell ``clock''. 
From the intensity of the correlation of this group with the cell-cycle, 
we conclude that this second clock is {\em not} strongly coupled to the 
cell cycle. Therefore this functional group is an independent oscillator. 
To elaborate this further we  have analyzed the frequency spectrum for 
the expression patterns of genes in this group and  for the cell-cycle 
regulated genes. Both spectral power distributions show a clear maximum at 
a frequency $\nu $, which differs significantly for the two groups. For the 
second clock we find $\nu = 4/7  \nu _{\mbox{cc}}$, where $\nu _{\mbox{cc}}$ 
is the frequency which maximizes the cell cycle power spectrum. The method 
used in \cite{ssziaebbf98} could therefore not identify the genes in this 
group as cyclic regulated.

A list of genes controlled by this cell-clock is available  at 
{\rm http://www.thep.lu.se/$\tilde{~}$sven}.
Unfortunately, the maximal time-span, for which experiments were done, 
contains only one complete cycle of this clock. It is therefore not possible 
to decide, if this is a continuous oscillator or a one-shot clock.
Some annotations which appear frequently for the genes found in this functional
group make it plausible, that this clock controls the transcription
of genes and the synthesis of proteins.

Next we want to show how the non-local part of the interaction in
(\ref{action}) influences the ordering of gene expression patterns. 
For each site $i$ in the list of expression patterns one can define 
an effective  prototype expression $P_i$, which is induced by 
{\em all} expression patterns in the data set via the energy.  
This prototype is the pattern which minimizes the energy, the solution 
$D_{i,t}$ of
\beq 
0 = \frac{\partial}{\partial D_{i,t}} S(\alpha, \beta), \qquad 
	t = 1 \ldots \tau .
\eeq
In general this optimal pattern will differ  from site to site. Hence it can 
follow a signal which is deformed from a pattern $A$ to a pattern $B$. 

To demonstrate this property we choose the cell-cycle regulated genes, 
where the signal ``activated'' travels through the system. Differently from
above, where the list of patterns was sorted with with respect to 
co-regulation, anti-correlated genes should {\em not} be grouped together in 
this case, because  presumably they belong to opposite parts of the cell 
cycle. We therefore choose $\gamma = 1$, when relabeling the expression 
patterns.  In figure \ref{Mitosis} the correlation 
coefficients for the  cell cycle regulated genes sorted in this way 
are displayed. We have used the analysis in \cite{ssziaebbf98} to assign
the genes to a specific part in the cell-cycle, the result is presented as
the annotation in the diagram. Obviously the data is correctly time-ordered
\footnote{Remember the list is cyclic, {\em i.e.} the last line is 
logically adjacent to the first line}. Note that the  algorithm does in 
no way explicitly refer to the time aspect in the data. In fact, the energy 
(\ref{action}) is invariant under the exchange of experiments, different 
time-points. This observation confirms the capability of the algorithm to 
identify a previously unknown signal-pathway hidden in the data. 
 
\turnmeoff{
\begin{figure}
\centerline{\psfig{file=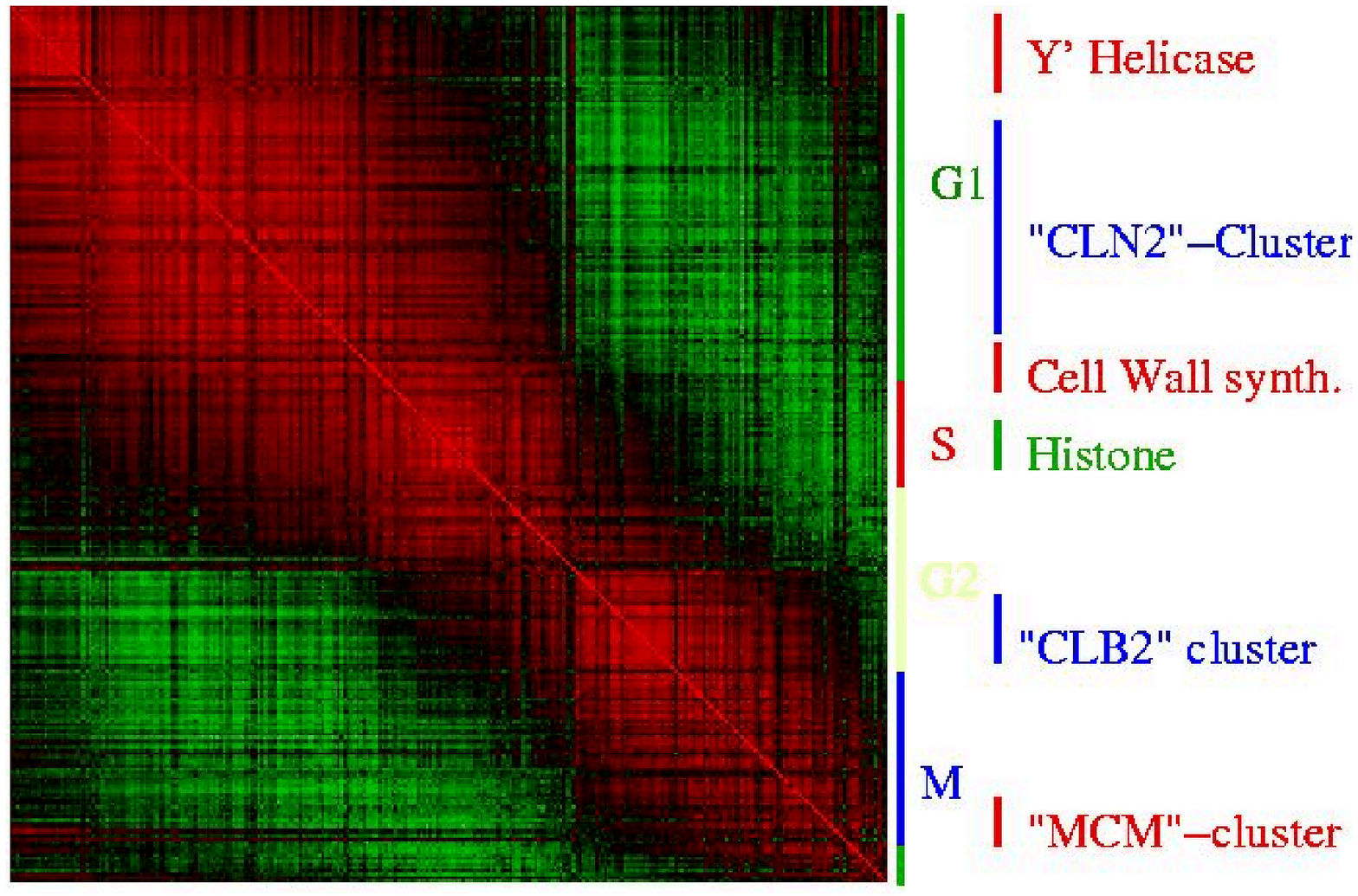,angle=0,height=6cm}}  
\caption{Correlation Matrix for the set of patterns annotated 
         ``Cell Cycle'' in figure (\ref{CorMat}) after resorting with
	$\gamma = 1$. Annotations following \cite{ssziaebbf98}.}   
\label{Mitosis}
\end{figure}
}
 
\section{Discussion}
Re-shuffling is efficient in finding functional groups in the expression data.
The philosophy of this algorithm  is considerably different from cluster 
algorithm, which compare each pattern $p$ locally
with some prototypic pattern for each cluster and finally assign $p$ 
to the cluster with the most similar prototype. In this way many of 
these algorithms do not use the information available about inter-cluster 
similarities. Re-shuffling is based on a global comparison with {\em all} 
patterns in  the ensemble. In this way it makes use of the information 
contained in sub-leading similarities as well. We want to emphasize 
that this method is not restricted to the analysis of time expression
data. It can also be used to detect patterns in static data to identify,
for example, genes which are responsible for a certain phenotype or
disease. 

Self organizing maps  use a non-local assignment of patterns to neurons.
Therefore they can reflect inter-cluster similarities. They
were used in \cite{tsmzkldlg99, TKWC99} to classify yeast gene expression 
patterns. With this method it is possible to identify the cell
cycle oscillation as a dominant motif in the expression data
\cite{tsmzkldlg99}. However, the neurons most active
for the corresponding patterns at different parts of the cell cycle
were not grouped in an obvious way. It was not possible to identify
these patterns as belonging to the same functional cycle. 

The re-shuffling method is able to extract this information.  
Without any prior knowledge it can identify the cell cycle regulated genes.
It is very interesting to observe  that the algorithm extracts 
time information from the data without actually referring to
the time aspect. This demonstrates that the pathways of signals can
be extracted from the data using this method. This aspect can be very
useful when analyzing functional groups which are not so well known
as the cell cycle. A further example of the algorithms power is the
identification of  a second independent clock in yeast. 

A general problem when analyzing expression data is noise.
When measurements are easily available, the usual way to reduce noise is
to increase the number $M$ of measurements until the noise level 
$\sigma \propto 1 / \sqrt{M}$ is small enough. Repeated measurements 
of the same system would also allow for a reliable estimation of the 
noise level. This knowledge is crucial when interpreting the
results of an analysis. However, gene expression measurements are quite 
costly in time and are usually not repeated. Therefore the methods used to 
analyze the data have to be relatively insensitive to noise. 
The re-shuffling algorithm is very robust in this respect  because 
the energy function (\ref{action}) used in the sorting procedure averages over
{\em all}  patterns in the data set. Many cluster algorithms and self 
organizing maps only average over a subset of the data when extracting 
a prototype for a cluster (or a neighborhood of a neuron). They are therefore
more sensitive with respect to noise. 

The visualization of the correlation matrix gives some insight into 
the connectivity of the underlying regulatory network. One may ask if it
is possible to learn the full  network from the data. The complexity of 
the model that can be inferred from the data is strictly limited by Shannons
theorem 
\beq
\frac{I_{\mbox{\tiny xmit}}}{N} = 
   \nu _{\mbox{\tiny max}} \log _2 \left ( 1 + \frac{S}{P} \right ).
\label{info}
\eeq
The signal $S$ -- the mRNA concentration in the cell-plasma --  is 
``transmitted'' over a  channel (the measurement process) with   
noise  $P$. With an estimated $15 \%$ noise-level and  the bandwidth 
given by the Nyquist frequency $\nu _{\mbox{\tiny max}} = 1/2$, the maximal
amount of information extractable from  one measurement of $N$ ORF's 
is less than $I_x = \frac{3 N}{2}$ bits. This is the theoretical upper
limit on the information content in the data. It  can only be reached if an 
optimal code is used. This is certainly not the case for the gene 
expression data, as it contains a lot of redundancy. The maximal complexity
of a model which can be inferred from the data is therefore considerably
smaller than estimated by equation (\ref{info}).  The amount of information
required to describe the connectivity of a possibly fully connected network
of $N$ nodes is $I_f \approx N (N-1) / 2 >> \tau I_{\mbox{\tiny xmit}}$,
much larger than 
available from the  $\tau $ measurements. Even if one restricts the maximal 
number of connections to $k$, the information $I_r \approx N k \log _2 N$
required to describe the connectivity of this network is larger than 
available already for small $k$. It seems therefore necessary to reduce
the number $N$ of nodes in the network. This can be done  by introducing 
collective nodes representing a whole subset of the original genes. 
The groups resulting from re-shuffling the data might be a good
starting point for this.          

\forget{
The  amount of information required to describe the connectivity of 
a possibly fully connected boolean network is 
$ I_{fb} = N ( N - 1) / 2 $, one needs at least $M \approx  N / 3.5$ 
Measurements. A boolean network with maximal $k$ connections per node requires 
$I_{bpc} = \frac{N k \log _2 N}{2}$ bits, $M = \frac{k \log _2 N}{3.5}$
Measurements. } 

In the future the algorithm can be modified to operate in a higher
dimensional index space. In this way a more refined representation
of structure found in the data is possible. It may also be useful 
to combine hierarchical clustering with this method, which can be 
used to freeze the orientation degree of freedom on each branching point
in the similarity tree.

\section{Acknowledgement}
I want to thank {\AA}.~Borg, C.~Peterson and M.~Ringn\'er for 
fruitful discussions. This work was supported by the 
Swedish Foundation for Strategic Research.

\turnmeon{
\section*{Figure captions}
{\sc Fig.~1.}\quad Correlation Matrix for the 800 most variant gene expression
         patterns. The annotation refers to the most frequent annotation
         found in the database for the genes in this group.

{\sc Fig.~2.}\quad Correlation Matrix for the set of patterns annotated 
         ``Cell Cycle'' in figure (\ref{CorMat}) after resorting with
 	 $\gamma = 1$. Annotations following \cite{ssziaebbf98}.   

\begin{figure}
\centerline{\includegraphics[angle=0,height=7cm,width=7cm]{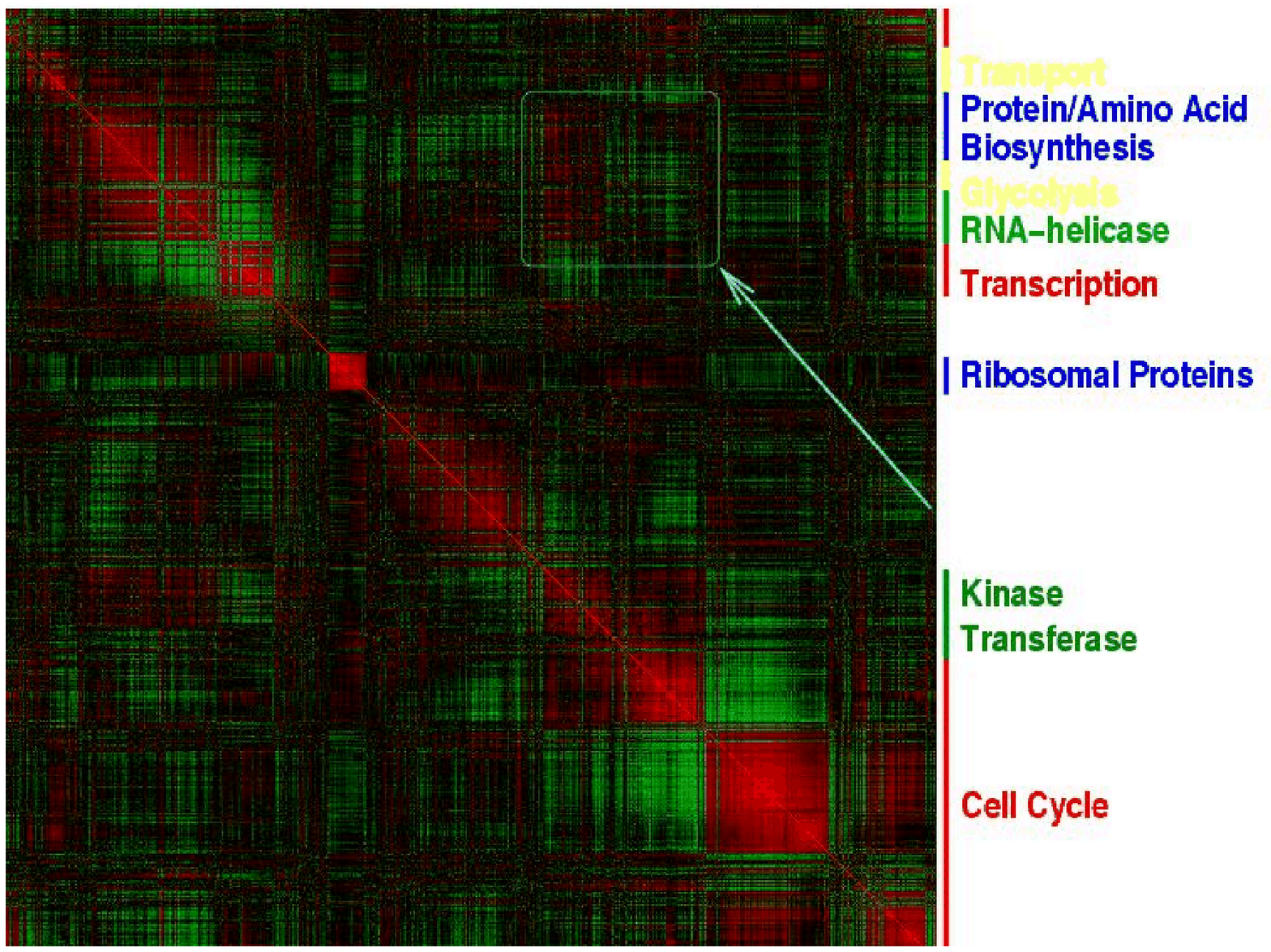}}
\vspace*{3mm}

\centerline{Fig. 1}
\label{CorMat}
\end{figure}
\newpage

\begin{figure}
\centerline{\includegraphics[angle=0,height=6cm,width=6cm]{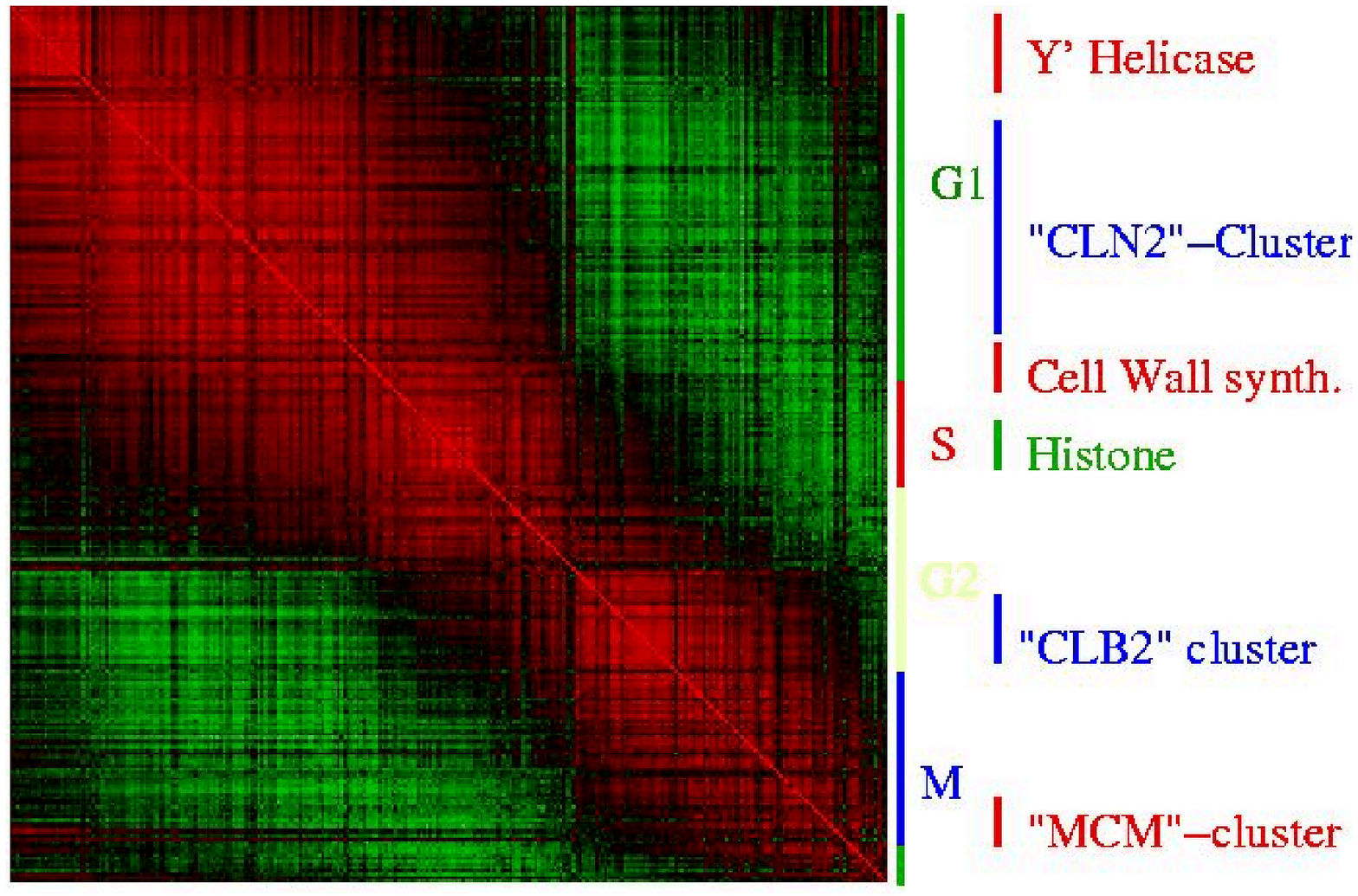}}
\vspace*{3mm}

\centerline{Fig. 2}
\label{Mitosis}
\end{figure}
}

\end{document}